\documentclass[12pt]{article}
\usepackage{amssymb}
\def\beq{\begin{equation}}
\def\eeq{\end{equation}}
\usepackage[all,2cell,dvips]{xy}

\def\IR{\relax{\rm I\kern -.18em R}}

\begin{document}
\title{Hamiltonian structure of an operator valued
extension of Super KdV equations }
\author{ \Large  A. Restuccia*, A. Sotomayor**}
\maketitle{\centerline{*Departamento de F\'{\i}sica}}
\maketitle{\centerline{Universidad de Antofagasta}
\maketitle{\centerline{*Departamento de F\'{\i}sica}}
\maketitle{\centerline{Universidad Sim\'on Bol\'{\i}var }

\maketitle{\centerline{**Departamento de Matem\'aticas}
\maketitle{\centerline{Universidad de Antofagasta}}
\maketitle{\centerline{e-mail: arestu@usb.ve,
adrian.sotomayor@uantof.cl }}
\begin{abstract}An extension of the super Korteweg-de Vries integrable system
in terms of operator valued functions is obtained. In particular
the extension contains the $N=1$ Super KdV and coupled systems
with functions valued on a symplectic space. We introduce a Miura
transformation for the extended system and obtain its hamiltonian
structure. We also obtain an extended Gardner transformation which
allows to find an infinite number of conserved quantities of the
extended system.\end{abstract}

\section{Introduction} The $N=1$ supersymmetric
extension of Korteweg-de Vries equation obtained in
\cite{Mathieu} is equivalent to the equations obtained in
\cite{Manin} by reduction from the super Kadomtsev-Petviashvili
hierarchy. The equations are described by a system of nonlinear
coupled partial differential equations for fields which take
values on the even and odd parts of a Grassmann algebra. The
system posses a sequence of infinite local as well as non-local
conserved quantities \cite{Mathieu,Mathieu1,Andrea,Adrian1}.
Later on, also supersymmetric extensions with more than one
supersymmetric generator were obtained
\cite{Mathieu2,Ivanov,Delduc1,Delduc2,Popowicz}.

Coupled partial differential equations in terms of real commuting
fields may be obtained from these supersymmetric models by
expanding the even and odd fields in terms of a basis of the
Grassmann algebra and separating the equations corresponding to
each generator. Families of new solutions \cite{Gao1,Gao2} to
super KdV were obtained using such bosonization procedure
\cite{Adrian}.

The bosonization approach becomes non-trivial when one considers
the hamiltonian formulation, the Poisson structure on the
unconstrained phase space and the Poisson structure on the
constrained phase space given by the Dirac brackets. In fact, the
formulation in terms of even variables introduces antisymmetric
Poisson brackets while the one in terms of odd variables
introduces symmetric Poisson brackets and the equivalence has to
be shown explicitly. In the case we will consider in this paper
both approaches becomes completely equivalent.

In this paper we will consider a very general algebraic structure
for the fields describing the nonlinear systems. The systems are
coupled partial differential equations extending the KdV
equation. In particular our KdV extensions include the
supersymmetric $N=1$ SKdV system as well as deformations of it.
It also describe systems whose odd part satisfies a symplectic
bracket structure.

We will obtain the hamiltonian structure of these extended KdV
systems. The Poisson structure on the constrained submanifold of
phase space will be given in terms of the Dirac brackets defined
for constrained systems \cite{Dirac,Nutku,Kentwell,Adrian2}. The
extended KdV systems we will consider have an infinite sequence of
conserved quantities which may be obtained via a Gardner
transformation as was done in \cite{Gardner,Adrian}.

\section{Extension of KdV equation, hamiltonian structure and infinite sequence of conserved quantities}
We consider an associative algebra of even and odd elements. The
even elements belong to a commutative algebra $ \mathcal{P}$ with
unit while the odd ones belong to $ \mathcal{Q}$ which satisfies
\beq
\begin{array}{ll}\mathcal{Q}\mathcal{P}\subset \mathcal{Q} \\ \left[\mathcal{Q},\mathcal{P}\right]=0
\\ \left[\mathcal{Q},\mathcal{Q}\right]\subset \mathcal{P},
\end{array}\label{eq1}\eeq and for any $q\in\mathcal{Q}$ there always
exists $\hat{q}\in\mathcal{Q}$ such that $
\left[q,\hat{q}\right]\neq0$.

We introduce a Lagrangian formulated in terms of fields $w$ and
$\eta$ valued on $ \mathcal{P}$ and $ \mathcal{Q}$ respectively.
The Lagrangian depends on a real parameter $ \lambda$. The
Lagrangian is
\beq\mathcal{L}=\frac{1}{2}\dot{w}w^\prime+\frac{\lambda}{2}\left[\dot{\eta},\eta\right]-\frac{1}{2}{(w^{\prime\prime})}^2
-\frac{1}{2}{(w^\prime)}^4-\frac{1}{2}\lambda^2{\left[\eta,\eta^\prime\right]}^2-\frac{\lambda}{2}\left[\eta^{\prime\prime},\eta^\prime\right]
-\frac{3}{2}\lambda{(w^\prime)}^2\left[\eta^\prime,\eta\right].\label{eq2}\eeq
We notice that if $\lambda$ is positive one may redefine
$\eta\rightarrow \lambda\eta$ and reduce the Lagrangian to the
case $\lambda=+1$. If $\lambda$ is negative one may reduce to the
case $\lambda=-1$.

The associated field equations are
\beq\dot{v}+v^{\prime\prime\prime}-6v^2v^\prime-3\lambda{\left(v\left[\eta^\prime,\eta\right]\right)}^\prime=0
\label{eq3} \eeq \beq \dot{\eta}+\eta^{\prime\prime\prime}-3
v^2\eta^\prime-3
vv^\prime\eta+\lambda\eta^\prime\left[\eta,\eta^\prime\right]+\frac{1}{2}\eta\lambda
\left[\eta,\eta^{\prime\prime}\right]=0\label{eq4}\eeq where
$v\equiv w^\prime$.

In particular if $ \mathcal{P}$ and $ \mathcal{Q}$ generates a
Grassmann algebra, the latter two terms on the second equations
are zero.

There is a hamiltonian and a corresponding Poisson structure
associated to this Lagrangian.

We denote by $p$ and $\mu$ the conjugate momenta associated to
$w$ and $\eta$ respectively.

We then obtain from the definition of $p$ and $\mu$ the following
primary constraints \beq \begin{array}{cc}\phi\equiv
p-\frac{1}{2}v=0 \\ \\ \psi\equiv \mu-\frac{\lambda}{2}\eta=0.
\end{array} \label{eq5}\eeq It turns out that these are the only
constraints of the theory. They are second class constraints. In
fact, they satisfy the following Poisson bracket relations
\begin{eqnarray*} & &\left\{\phi(x),\phi(y)\right\}_{PB}=-\partial_x\delta(x,y) \\ & & \left\{\phi(x),\psi(y)\right\}_{PB}=0
\\ & & \left\{\psi(x),\psi(y)\right\}_{PB}=-\frac{1}{\lambda}\delta(x,y).
 \end{eqnarray*} The hamiltonian density associated to the Lagrangian (2) may
 be obtained via a Legendre transformation and is expressed as
 \beq \mathcal{H}=\frac{1}{2}{(v^\prime)}^2+\frac{1}{2}v^4+\frac{1}{2}\lambda^2{\left[\eta,\eta^\prime\right]}^2
 +\frac{1}{2}\lambda\left[\eta^{\prime\prime},\eta^\prime\right]+\frac{3}{2}\lambda
 v^2\left[\eta^\prime,\eta\right]\label{eq6}\eeq subject to the
 constraints $\phi=0,\psi=0.$

 The canonical field equations \beq \begin{array}{ll}\dot{v}=\left\{v,H\right\}_{DB} \\ \dot{\eta}=
 \left\{\eta,H\right\}_{DB}  \end{array}  \label{eq7}\eeq where
 $H={\left\langle\mathcal{H}\right\rangle}_x$ is the integral on $
 \mathbb{R}$, exactly agree with the Lagrangian field equations
 (3),(4) as it should be.

 We notice that if $ \mathcal{P}$ and $ \mathcal{Q}$ generate a
 Grassmann algebra $H$ reduces to the hamiltonian of the $N=1$
 supersymmetric KdV equations.

 In fact, in the case of a Grassmann algebra the term
 ${\left[\eta,\eta^\prime\right]}^2$ in $ \mathcal{H}$ becomes
 zero and we may perform a Miura transformation
 \beq \begin{array}{ll}u=v^\prime+v^2-\lambda\left[\eta,\eta^\prime\right] \\ \xi=\eta^\prime+v\eta \end{array}
 \label{eq8}\eeq to obtain
 \[\mathcal{H}=\frac{1}{2}u^2+\frac{1}{2}\lambda\left[\xi^\prime,\xi\right]=\frac{1}{2}u^2+\lambda\xi^\prime\xi.\]
 The canonical equations then reduce to the system
 \beq\begin{array}{ll}u_t= -u^{\prime\prime\prime}+6 uu^\prime-6\lambda
\xi\xi^{\prime\prime}\\
\xi_t=-\xi^{\prime\prime\prime}+3{(u\xi)}^\prime,\end{array}\label{eq9}
\eeq which is invariant under the supersymmetric transformation
with odd parameter $\epsilon$, \beq \begin{array}{ll}\delta_\epsilon u=2\epsilon\lambda\xi^\prime\\
\delta_\epsilon\xi=\epsilon u.\end{array} \label{eq10}\eeq (9) is
a parametric Susy KdV equation, for $ \lambda=1$ it gives the
$N=1$ Super KdV equation.

If instead of considering a Grassmann algebra, we consider an
associative algebra satisfying in addition to (1) the condition $
\mathcal{Q}\mathcal{Q}\subset\mathcal{P}$, which implies $
[\mathcal{Q},\mathcal{Q}]\subset\mathcal{P},$ then the
hamiltonian density reduces to
\[\mathcal{H}=\frac{1}{2}{(v^\prime)}^2+\frac{1}{2}v^4+\frac{1}{2}\lambda\left[\eta^{\prime\prime},\eta^\prime\right]+\frac{3}{2}\lambda
 v^2\left[\eta^\prime,\eta\right].\] After using the generalized
 Miura transformation given by (8) we obtain
 \[\mathcal{H}=\frac{1}{2}u^2+\frac{\lambda}{2}[\xi^\prime,\xi]\] and
 the canonical field equations \beq\begin{array}{ll}u_t= -u^{\prime\prime\prime}+6 uu^\prime+3\lambda
[\xi^{\prime\prime},\xi]\\
\xi_t=-\xi^{\prime\prime\prime}+3{(u\xi)}^\prime,\end{array}\label{eq11}\eeq
which are invariant under the transformations
\beq\begin{array}{ll}\delta_\epsilon u= \lambda[\epsilon,\xi^\prime]\\
\delta_\epsilon\xi=\epsilon u.\end{array}\label{eq12}\eeq
Moreover, the system (11) under assumption (1) has an infinite
sequence of local conserved quantities for any value of
$\lambda$. This property may be proven by using a Gardner
transformation as was done for the case $\lambda=1$ in
\cite{Adrian}. In fact, we have

\beq \begin{array}{ll}
z_t={(-z^{\prime\prime}+3z^2+3\lambda\left[\sigma^\prime,\sigma\right])}^\prime+\epsilon^2{(2z^3+3\lambda
z[\sigma^\prime,\sigma])}^\prime
\\ \sigma_t={(-\sigma^{\prime\prime}+3z\sigma)}^\prime+\epsilon^2
3(z^2\sigma^\prime+z z^\prime
\sigma+\lambda\sigma^\prime[\sigma^\prime,\sigma]),\label{eq13}\end{array}\eeq

\beq \begin{array}{ll} u = z+\epsilon z^\prime+\epsilon^2(z^2+\lambda[\sigma^\prime,\sigma])\\
 \xi = \sigma+\epsilon \sigma^\prime
 +\epsilon^2z\sigma,\end{array}
\eeq where (13) and (14) are the Gardner equations and associated
Gardner transformations respectively.

After simplifying by crossing out derivatives in $x$ and using
the inverse Gardner transformation, the first four nontrivial
conserved quantities for the operator-extended KdV system (11)
are:

\begin{eqnarray}
H_0&=&\int udx \nonumber \\ H_2&=&\int
\left(u^2+\lambda[\xi^\prime, \xi]\right)dx \nonumber \\
H_4&=&\int\left(2u^3+{(
u^\prime)}^2+4\lambda u\left[\xi^\prime,\xi\right]+\lambda\left[\xi^{\prime\prime},\xi^\prime\right]\right)dx  \\
H_6&=&\int
\left(5u^4+10u{(u^\prime)}^2+{(u^{\prime\prime})}^2+15\lambda
u^2\left[\xi^\prime,\xi\right]-2\lambda
u\left[\xi^{\prime\prime},\xi^\prime\right]\right. \nonumber
\\ &-& \left.8\lambda u\left[\xi^{\prime\prime\prime},\xi\right]+3\lambda^2{\left[\xi^\prime,\xi\right]}^2+\lambda\left[\xi^{\prime\prime\prime},
\xi^{\prime\prime}\right]\right)dx. \nonumber
\end{eqnarray}

\section{Conclusions} We introduced an operatorial extension for
Korteweg-de Vries equation which contains as particular cases
several systems with the property of having a sequence of
infinite local conserved quantities. In particular it contains
the $N=1$ super KdV system. We obtained the hamiltonian structure
of such extension. The existence of an infinite sequence of
conserved quantities for the operatorial extension was shown
using a generalized Gardner transformation.
 $\bigskip$

\textbf{Acknowledgments}

A. R. and A. S. are partially supported by Project Fondecyt
1121103, Chile.

\end{document}